\documentstyle[aps,epsfig]{revtex}
\begin{document} 

\title{$SU(3)$ symmetry breaking in lower $fp$-shell nuclei}
\author{V. G. Gueorguiev \footnote{Corresponding author. Email address:
vesselin@phys.lsu.edu}, J. P. Draayer and C. W. Johnson} 
\address{Department of Physics and Astronomy,\\
Louisiana State University,\\
Baton Rouge, LA 70803\\}
\date{DOI: 10.1103/PhysRevC63.014318}

\maketitle 

\vskip 1cm

\begin{abstract} 

Results of shell-model calculations for lower $fp$-shell nuclei show that
$SU(3)$ symmetry breaking in this region is driven by the single-particle
spin-orbit splitting. However, even though states of the yrast band exhibit
$SU(3)$ symmetry breaking, the results also show that the yrast band $B(E2)$
values are insensitive to this fragmentation of the $SU(3)$ symmetry;
specifically, the quadrupole collectivity as measured by $B(E2)$ transition
strengths between low lying members of the yrast band remain high even though
$SU(3)$ appears to be broken. Results for $^{44,46,48}Ti$ and $^{48}Cr$ using
the Kuo-Brown-3 two-body interaction are given to illustrate these observations.

\end{abstract}


\vskip 1cm


\section{Introduction}

$SU(3)$ is a special algebraic structure because it is the compact symmetry
group of the three-dimensional isotropic harmonic oscillator \cite{SU(3) and 3D
HO} which is a good first-order approximation to any attractive potential. This
applies in nuclear physics and is the underpinning to the Elliott $SU(3)$ model
\cite{Elliott's model}. In the latter case the highest symmetry group is
$SU(\kappa\Omega)$, where $\Omega$ denotes the degeneracy of the spatial
degrees of freedom and $\kappa$ counts the number of internal degrees of
freedom (for example, $\kappa=2$ for identical spin $\frac{1}{2}$ particles and
$\kappa=4$ for a spin-isospin system). $SU(3)$ enters in this picture through a
reduction of $SU(\kappa\Omega)$ into spatial [$SU(\Omega)$] and spin or
spin-isospin degrees of freedom [$SU(\kappa)$], namely, $SU(\kappa\Omega)
\supset SU(\Omega) \otimes SU(\kappa )$, followed by a reduction of the spatial
degrees of freedom through $SU(3)$ to $SO(3)$; that is, $SU(\Omega)\supset
SU(3) \supset SO(3)$. Interactions that are not functions of the $SU(3)$
generators induce $SU(3)$ symmetry breaking. The spin-orbit interaction, which
is needed for a correct description of shell and subshell closures \cite{ls -
correct shell closures}, is an example of a one-body $SU(3)$ symmetry breaking
interaction while the pairing interaction, which is required for a correct
description of binding energies \cite{pairing interaction for the g.s.}, is an
example of a two-body $SU(3)$ symmetry breaking interaction. 

It is well known that $SU(3)$ is a very useful symmetry in the lower $sd$-shell
\cite{Elliott's model}. This is most easily understood by noting that the
leading irreducible representation (irrep) of $SU(3)$ normally suffices to
achieve a good description of the low-lying eigenstates of these nuclei. In the
lower $fp$-shell, however, leading $SU(3)$ irreps do not provide satisfactory
results for low-lying eigenstates. Beyond the $fp$-shell, the concept of
pseudospin symmetry \cite{pseudo SU(3) symmetry} allows one to identify
another so-called pseudo $SU(3)$ structure that again yields a good description
of low-lying eigenstates of strongly deformed nuclei \cite{Pseudo SU(3) in the
strongly deformed nuclei}. Questions that remain regarding the lower $fp$-shell
are: What parts of the interaction are responsible for the $SU(3)$ symmetry
breaking? Is it the one-body part, the two-body part, or a combination of these
two? And if it is a combination, to what extent does each interaction
contribute to $SU(3)$ symmetry breaking? Also, what is the effect of the
$SU(3)$ symmetry breaking on the electromagnetic transition rates? Enhanced
$B(E2)$ transition rates \cite{B(E2) expressions} are normally considered to be
a good indicator of quadrupole collectivity and the $SU(3)$ structure of the
corresponding initial and final states. It has been suggested that strong
$B(E2)$ values may survive an ``adiabatic'' mixing of $SU(3)$ irreps due to
quasi-$SU(3)$ dynamical symmetry \cite{Adiabatic mixing}. A signature for
this type of mixing is $B(E2)$ values that are similar to those obtained when
the $SU(3)$ symmetry is good. Is the $SU(3)$ symmetry breaking in the lower
$fp$-shell adiabatic?

In this paper we show for lower $fp$-shell nuclei that whereas the spin-orbit
interaction is the primary driver of $SU(3)$ symmetry breaking the $B(E2)$
values between the first few yrast states remain strong, signaling an adiabatic
mixing of $SU(3)$ irreps. The realistic monopole-corrected Kuo-Brown-3 two-body
interaction \cite{KB3 interaction} is used in calculations for $^{44,46,48}Ti$
and $^{48}Cr$ with single-particle energies corresponding to realistic
spin-orbit splitting. The spectrum of the second-order Casimir operator $C_{2}$
of $SU(3)$ is used as a measure for gauging the $SU(3)$ fragmentation along the
yrast band of these nuclei. The results show that the spin-orbit splitting is
the primary cause for $SU(3)$ symmetry breaking; the leading $SU(3)$ irrep
regains its importance as the spin-orbit splitting is turned off. A similar
recovery of the $SU(3)$ symmetry has been reported in the case of $^{44}Ti$
with degenerate $f_{\frac{7}{2}}-p_{\frac{3}{2}}$ shells \cite{SU(3) and
Scissors in sd-shell}.

To fix the notation, in the following section a parameterization of the
Hamiltonian in terms of one-body spin-orbit and orbit-orbit single-particle
interactions, as well as a general two-body interactions, is given. In our
applications of the theory, the realistic Kuo-Brown-3 interaction is chosen for
the two-body interaction \cite{KB3 interaction}. Computational methods used in
the analyses are discussed in the third section. This is followed by
characteristic results for $^{44}Ti$, $^{46}Ti$, $^{48}Ti$, and $^{48}Cr$ in
the fourth section. A conclusion that recaps outcomes is given in the fifth and
final section. 

\section{Interaction Hamiltonian}

The one- plus two-body Hamiltonian is used in standard second-quantized form: 

\[
H=\sum_{i}\varepsilon _{i}a_{i}^{+}a_{i}+ \frac{1}{4}
\sum_{i,j,k,l}V_{kl,ij}a_{i}^{+}a_{j}^{+}a_{k}a_{l}.
\]

\noindent
The summation indexes range over the single-particle levels included in the
model space. We only consider levels of the $fp$-shell which have the following
radial $(n)$, orbital $(l)$ and total angular momentum $(j)$ quantum numbers:
$nl_{j}=\left\{ 0f_{\frac{7}{2}}, 0f_{\frac{5}{2}}, 1p_{\frac{3}{2}},
1p_{\frac{1}{2}}\right\}$. In what follows the radial quantum number $(n)$ is
dropped since the $l_{j}$ labels provide a unique labelling scheme for
single-shell applications. It is common practice to replace the four
single-particle energies $\varepsilon _{i}$ in the $fp$-shell by the $l^{2}$
and $l\cdot s$ interactions: $\sum_{i}\varepsilon _{i}a_{i}^{+}a_{i}\rightarrow
\epsilon (n_{i}-\alpha _{i}l_{i}\cdot s_{i}-\beta _{i}l_{i}^{2})$, where
$\epsilon$ is the average binding energy per valence particle, $n_{i}$ counts
the total number of valence particles, and $\alpha$ and $\beta$ are
dimensionless parameters giving the interaction strength of the $l^{2}$ and
$l\cdot s$ terms. For realistic single-particle energies used in the KB3
interaction (\ref{KB3 spe}), these parameters are $\epsilon=2.6$ $MeV,$
$\beta=0.0096,$ $\alpha _{p}=1.3333,$ and $\alpha _{f}=1.7143.$ The small value
of $\beta$ signals small $l^{2}$ splitting (\ref{KB3p-f spe}).

A significant part of the two-body interaction, $V_{kl,ij}$, maps onto the
quadrupole-quadrupole, $Q\cdot Q$, and the pairing, $P$, interactions. Since
$Q\cdot Q$ can be written in terms of $SU(3)$ generators, it induces no $SU(3)$
breaking and hence serves to re-enforce the importance of the Elliott model
\cite {Elliott's model}, when the pairing interaction mixes different $SU(3)$
irreps. In this analysis the two-body part of the Hamiltonian, $V_{kl,ij}$, is
fixed by the Kuo-Brown-3 (KB3) interaction matrix elements while the
single-particle energies, $\varepsilon _{i}$, are changed as described below.

The following single-particle energies are normally used with the KB3
interaction \cite{KB3 interaction}:

\begin{eqnarray} 
{\rm KB3\quad [MeV]} &:&
\varepsilon_{p_{\frac{1}{2}}}=4,\quad \varepsilon _{p_{\frac{3}{2}}}=2,
\label{KB3 spe} \\ &&\left. \varepsilon _{f_{_{\frac{5}{2}}}}=6,\quad
\varepsilon _{f_{\frac{7}{ 2}}}=0\right. \nonumber 
\end{eqnarray} 

For the purposes of the current study, it is important to know the
single-particle centroids of the $p$ and $f$ shells. For example, the energy
centroid of the $p$ shell is given by:

\[ 
\varepsilon _{p}= \frac{ \varepsilon _{p_{\frac{1}{2}}} \dim(p_{\frac{1}{2}})+
\varepsilon _{p_{\frac{3}{2}}} \dim (p_{\frac{3}{2}})} {\dim(p_{\frac{1}{2}})+
\dim (p_{\frac{3}{2}})}. 
\]

In what follows, we label by $KB3p\_f$ that Hamiltonian which uses the KB3
two-body interaction with single-particle $p$- and $f$-shell energies set to
their centroid values:

\begin{eqnarray}
{\rm KB3}p\_f\quad [MeV] &:&
\varepsilon_{p_{\frac{1}{2}}}=\varepsilon _{p_{ \frac{3}{2}}}=2.6670 
\label{KB3p-f spe} \\ && \left. 
\varepsilon _{f_{_{\frac{5}{2}}}}=\varepsilon_{f_{\frac{7}{2}}}=2.5710\right.
\nonumber 
\end{eqnarray}

We use $KB3pf$ for the case when the single-particle energies are set to their
overall average: 

\begin{equation}
{\rm KB3}pf\quad [MeV]:\quad\varepsilon _{p}=\varepsilon_{f}=2.6 
\label{KB3pf spe}
\end{equation} 

Due to the near degeneracy of the single-particle energies of the $KB3p\_f$
interaction (\ref{KB3p-f spe}), the results for the $KB3pf$ case are very
similar to those for $KB3p\_f$.

\section{Computational procedures}

The computational procedures and tools used in the analysis are described in
this section. In brief, the Hamiltonian and other matrices are calculated using
an $m$-scheme shell model code \cite{m-scheme shell model codes} while the
eigenvectors and eigenvalues are obtained by means of the Lanczos algorithm
\cite{Lanczos algorithm}. All the calculations are done in the full $fp$ -shell
model space.

First, the Hamiltonian $H$ for each interaction ($KB3$ (\ref{KB3 spe}),
$KB3p\_f$ (\ref{KB3p-f spe}), and $KB3pf$ (\ref{KB3pf spe})) is generated. Then
the eigenvalues and eigenvectors are calculated and the yrast states
identified. Next, the matrix for the second order Casimir operator of $SU(3)$,
namely $C_{2}=\frac{1}{4}(3L^{2}+Q\cdot Q)$, is generated using the shell model
code and a moments method \cite{Moments method} is used to diagonalize the
$C_{2}$ matrix by starting the Lanczos procedure with specific eigenvectors of
$H$ for which an $SU(3)$ decomposition is desired. Finally, $B(E2)$ values in
$e^{2}fm^{4}$ units are calculated from one-body densities using Siegert's
theorem with a typical value for the effective charge \cite{effective charges},
$q_{eff}$ = 0.5, so $e_p = (1+q_{eff})e = 1.5e$ and $e_n = (q_{eff})e = 0.5e$.

Although the used procedure can generate the spectral decomposition of a state
in terms of the eigenvectors of $C_{2}$ of $SU(3)$, this alone is not 
sufficient to uniquely determine all irreducible representation (irrep) labels 
$\lambda$ and $\mu$ of $SU(3)$. For example, $C_{2}$ has the same eigenvalue
for the $(\lambda,\mu)$ and $(\mu,\lambda)$ irreps. Nevertheless, since for the
first few leading irreps (largest $C_{2}$ values) the $\lambda$ and $\mu$
values can be uniquely determined \cite{Tabels of SU(N) to SU(3)} this
procedure suffices for our study.

Usually, when considering full-space calculations, a balance between computer
time and accuracy has to be considered. While the Lanczos algorithm
\cite{Lanczos algorithm} is known to yield a good approximation for the lowest
or highest eigenvalues and eigenvectors, it normally does a relatively poor job
for intermediate states. This means, for example, that higher states, in
particular high total angular momentum states, may be poorly represented or, in
a worst case scenario, not show up at all when these states are close to or
beyond the truncation edge of the chosen submatrix. An obvious way to maintain
a good approximation is to run the code for each $M_{J}$ value, that is, $
M_{J}=0,2,4,6$\ldots. However, this might be a very time consuming process, but
nonetheless one which could be reduced significantly if only a few
$M_{J}$ values are used for each run. For the calculations of this study, we
used $M_{J}=0,6,10,$ and $14$. To maintain high confidence in the approximation
of the intermediate states which have $J=2,4,8,12,...$ we required that they be
within the first half of all the states produced. The code was set up to output
$29$ states. A further verification on the accuracy of the procedure is whether
the energies of the same state calculated using different $M_{J}$ runs are
close to one another. For example, as a consistency check the energy of the
lowest $J=6$ state in the $M_{J}=0$ run was compared to the energy of the same
state obtained from the $M_{J}=6$ run. 

\section{Results}

Results for the $SU(3)$ content of yrast states and their $B(E2)$ values for
representative $fp$-shell nuclei are reported in this section. We focus on $
^{44}Ti$, $^{46}Ti$, $^{48}Ti$, and $^{48}Cr$ because these are $fp$-shell
equivalents of $^{20}Ne$, $^{22}Ne$, $^{24}Ne$, and $^{24}Mg$, respectively,
which are known to be good $SU(3)$ $sd$-shell nuclei. Furthermore, data on
these nuclei are readily available from the National Nuclear Data Center (NNDC)
\cite{NNDC} and full $fp$-shell calculations are feasible \cite{full fp shell}.
The model dimensionalities for full-space calculations increase very rapidly
when approaching the mid-shell region; those for the cases considered here are
given in Table \ref{space dimensions}.

\begin{table}
\begin{tabular}{rrrrrrrrr} 
Nucleus && $M_{J}=0$ && $M_{J}=6$ && $M_{J}=10$ && $M_{J}=14$ \\ 
\hline
$^{44}Ti$ && 1080 && 514 && 30 && --- \\ 
$^{46}Ti$ && 43630 && 32297 && 4693 && 134 \\
$^{48}Ti$ && 317972 && 278610 && 57876 && 3846 \\ 
$^{48}Cr$ && 492724 && 451857 && 104658 && 8997 \\ 
\end{tabular}
\caption{Space dimensions for $m$-scheme calculations in full
$fp$-shell model space. The computer code uses even parity and even isospin
basis states with no restrictions on the total angular momentum $J$ except for
$M_{J}=0$ case where the computer code selects only states with even $J$
values.} 
\label{space dimensions}
\end{table}

\begin{figure}[tbp] 
\centerline{\hbox{
\epsfig{figure=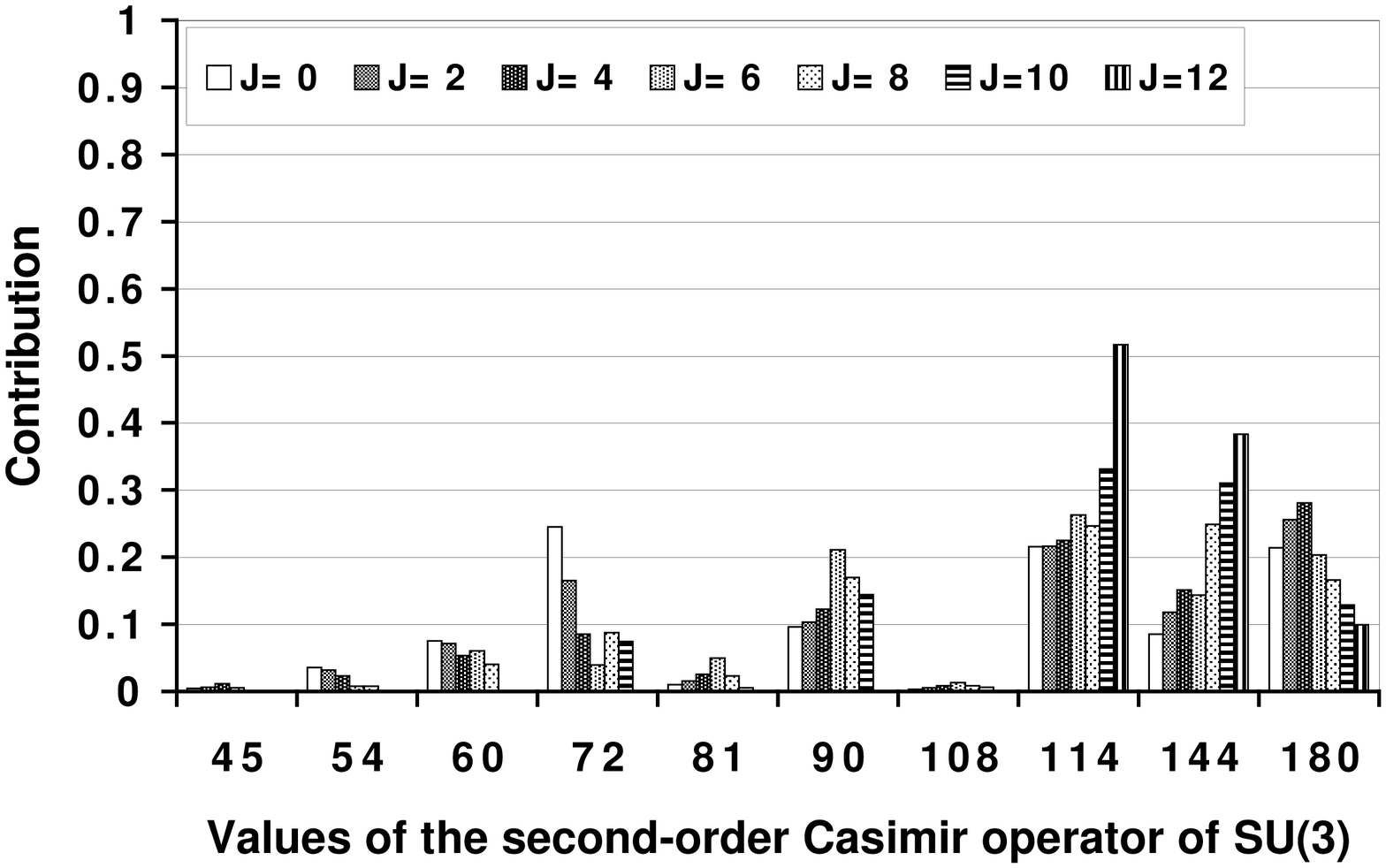,width=8cm,height=6cm}
}}
\caption{Strength
distribution of $C_{2}$ of $SU(3)$ in yrast states of $ ^{44}Ti$ for realistic
single particle energies with Kuo-Brown-3 two body interaction ($KB3$).}
\label{C2 of SU(3) for KB3 in Ti44}
\end{figure}

In the following, we use four different graphic representations to illustrate
our results. The first set, Figs.\ref{C2 of SU(3) for KB3 in Ti44} and \ref{C2
of SU(3) for KB3p_f in Ti44}, demonstrates the recovery of the $ SU(3)$
symmetry as the single-particle spin-orbit interaction is turned off, that is,
in going from the $KB3$ to the $KB3p\_f$ interaction. Corresponding results
for the $KB3pf$ interaction are not given since they are similar to the
$KB3p\_f$ results. In each graph, $C_{2}$ values of $SU(3)$ are given on the
horizontal axis with the contribution of each $SU(3)$ state on the vertical
axis. The bars within each cluster are contributions to the yrast states
starting with the ground state ($J=0$) on the left. Hence the second bar in
each cluster is for the $J=2$ yrast state, etc. 

\begin{figure}[tbp] 
\centerline{\hbox{
\epsfig{figure=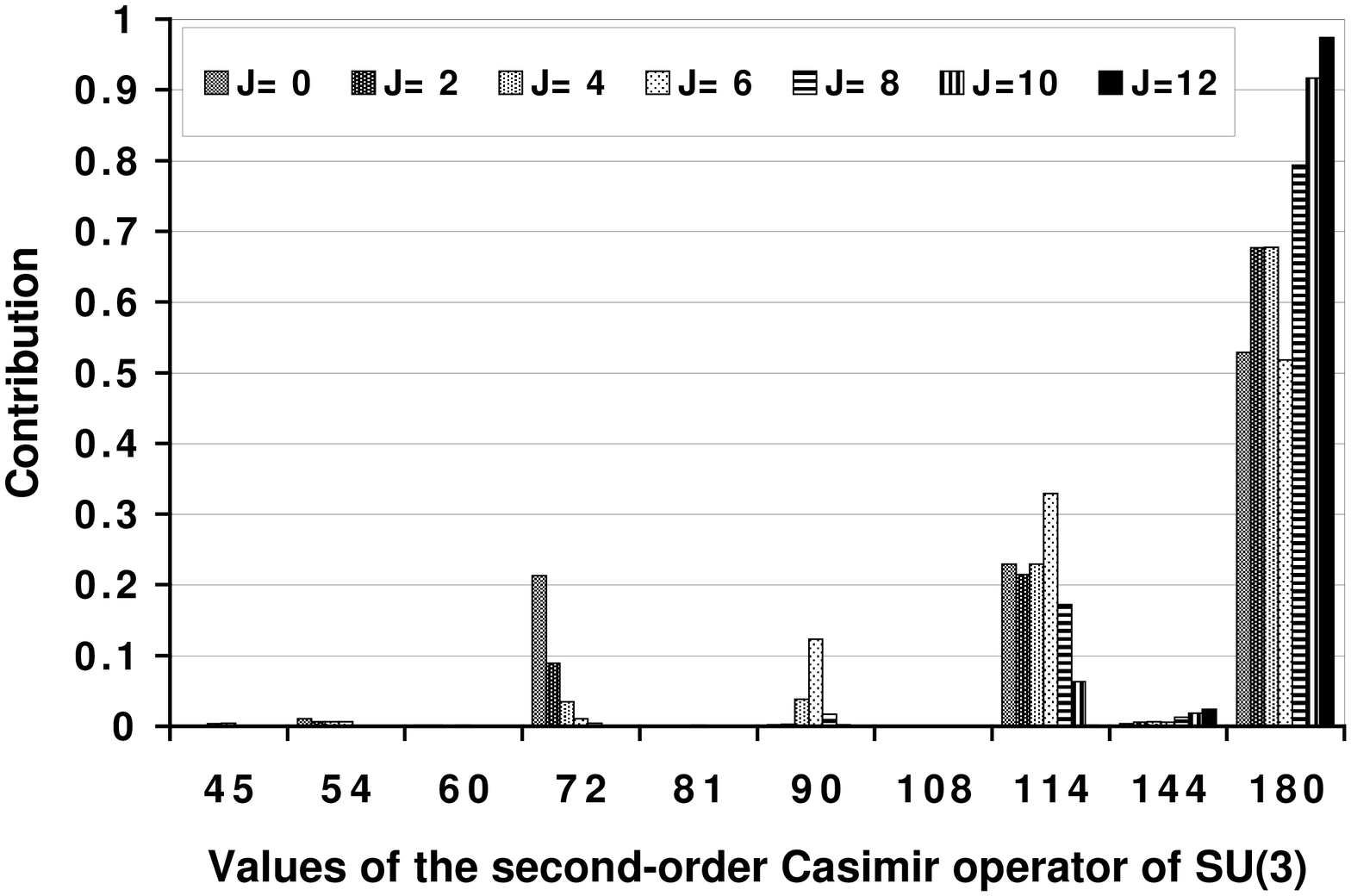,width=8cm,height=6cm}
}}
\caption{Strength distribution of $C_{2}$ of $SU(3)$ in yrast states of
$^{44}Ti$ for degenerate single particle energies with Kuo-Brown-3 two body
interaction ($KB3p\_f$).} 
\label{C2 of SU(3) for KB3p_f in Ti44}
\end{figure}

We have chosen $^{44}Ti$ for an in-depth consideration of the fragmentation of
the $C_{2}$ strength in yrast states. The results on the nondegenerate $KB3$
interaction are shown in Fig.{\ref{C2 of SU(3) for KB3 in Ti44}. In this case
the highest contribution (biggest bar) is more than $50\%$ which corresponds to
a $C_{2}$ value of 114 for the $J=12$ state. The $C_{2}=114$ value is for
$(\lambda ,\mu )=(8,2)$ which is two $SU(3)$ irreps down from the leading one,
$ (\lambda ,\mu )=(12,0)$ with $C_{2}=180$. The leading irrep only contributes
about 10$\%$ to the $J=12$ yrast state. The contribution of the next to the
leading irrep, $C_{2}=144$ for $(\lambda ,\mu )=(10,1)$, is slightly less than
40$\%$. Thus, for all practical purposes, the first three irreps determine the
structure of the $J=12$ yrast state. This illustrates that the high total
angular momentum $J$ states are composed of only the first few $ SU(3)$ irreps.
This is easily understood because high $J$ values require high orbital angular
momentum $L$ which are only present in $SU(3)$ irreps with large $C_{2}$
values. The high $J$ states may therefore be considered to be states with good
$SU(3)$ symmetry. However, this is not the case with the ground state of
$^{44}Ti$ which has very important contributions from states with $C_{2}$
values 60, 72, 90, 114, 144, and 180 with respective percentages, 7.5,
25, 10, 21, 8, and 21$\%$. This shows that the leading irrep is
not the biggest contributor to the $J=0$ ground state; there are two other
contributors with about $20\%$, the third ($C_{2}=114$) and seventh
($C_{2}=72$) $SU(3)$ irrep. }

When the spin-orbit interaction is turned off, which yields nearly degenerate
single-particle energies since the single-particle orbit-orbit splitting is
small, one has the $KB3p\_f$ interaction and in this case the structure of the
yrast states changes dramatically, as shown in Fig.\ref{C2 of SU(3) for KB3p_f
in Ti44}. From Fig.\ref{C2 of SU(3) for KB3p_f in Ti44} one can see that the
leading irrep plays a dominant role as its contribution is now more then $50\%$
of every yrast state. As in the previous case, the high total angular momentum
$J$ states have the biggest contributions from the leading irrep, for example,
more than 97$\%$ for $ J=12$, 91$\%$ for $J=10$, and 80$\%$ for $J=8$. The
ground state is composed of few irreps with $C_{2}$ values 72, 114, and 180,
but in this case the leading irrep with $C_{2}=180$ makes up more than 52$\%$
of the total with the other two most important irreps contributing 21$\%$ [$
C_{2}=72$, $(\lambda ,\mu )=(4,4)]$ and 23$\%$ [$C_{2}=114$, $(\lambda ,\mu
)=(8,2)].$

\begin{figure}[tbp]
\centerline{\hbox{
\epsfig{figure=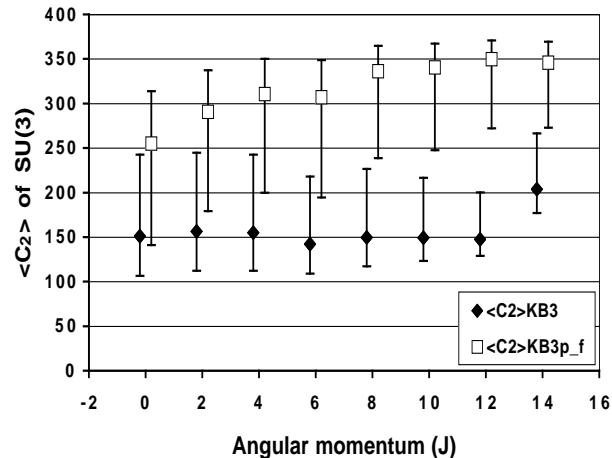,width=8cm,height=6cm}
}}
\caption{Average $C_{2}$ values for $KB3$ and $KB3p\_f$ interactions in
$^{48}Ti$.}
\label{<C2> for KB3 and KB3p_f in Ti48}
\end{figure}

An alternative way to show these results is given in Fig.\ref{<C2> for KB3 and
KB3p_f in Ti44} and Fig.\ref{<C2> for KB3 and KB3p_f in Ti48}. These figures
show the centroid, width, and skewness of the $C_{2}$ distributions. The $J$
values are plotted on the horizontal axis with the centroids given on the
vertical axis. The width of the distribution is indicated by the length of the
error bars which is just the rms deviation, $\Delta
C_{2}=\sqrt{\left\langle\left( C_{2}-\left\langle C_{2}\right\rangle \right)
^{2}\right\rangle }$, from the average value of the second-order Casimir
operator $\left\langle C_{2}\right\rangle$. The third central moment, $\delta
C_{2}=\sqrt[ 3 ]{\left\langle \left( C_{2}-\left\langle C_{2}\right\rangle
\right) ^{3}\right\rangle }$, which measures the asymmetry, is indicated by the
length of the error bar above, $\Delta C_{2}+\frac{\delta C_{2}}{2}$, and
below, $\Delta C_{2}-\frac{\delta C_{2}}{2}$, the average value. 

\begin{figure}[tbp]
\centerline{\hbox{
\epsfig{figure=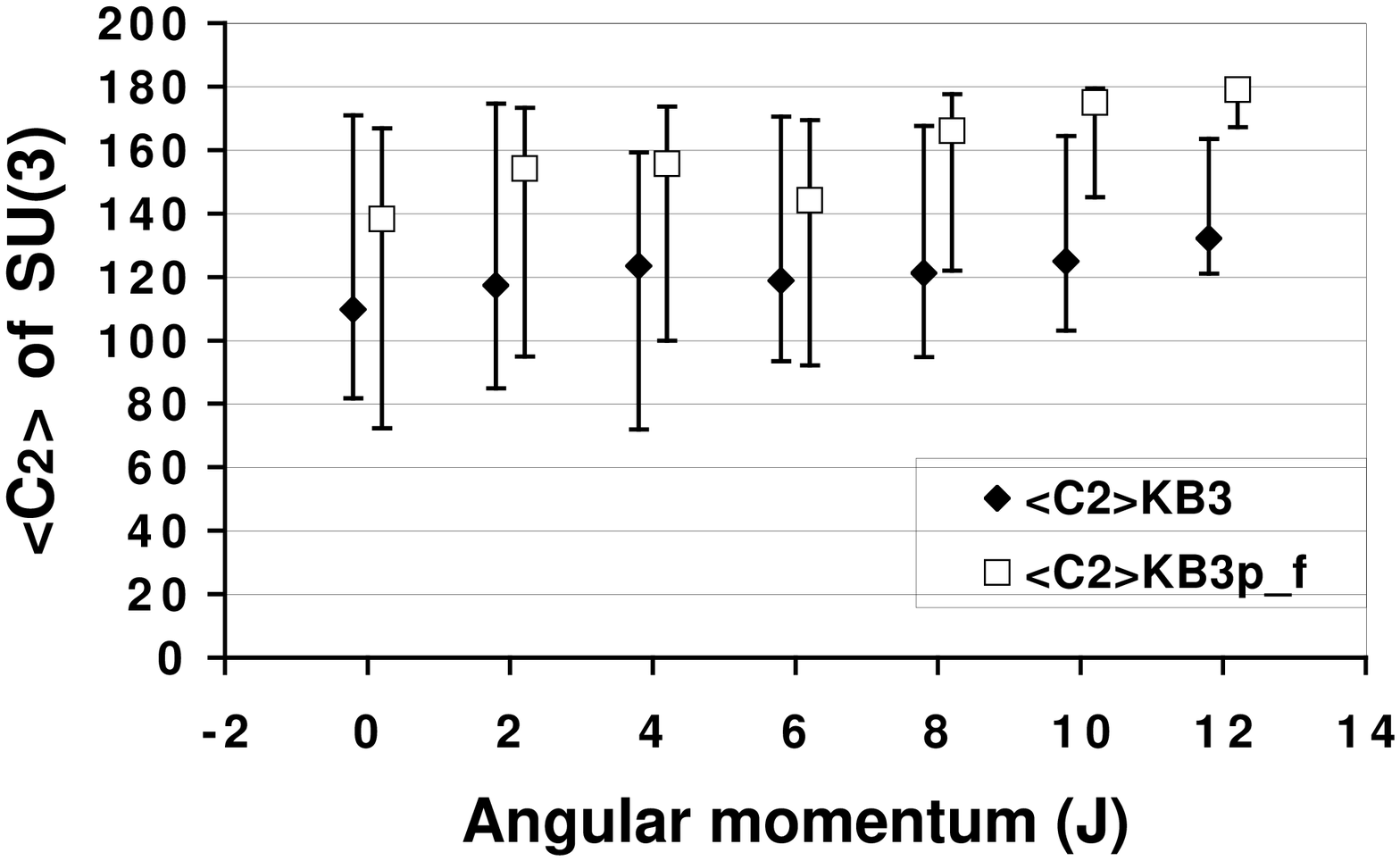,width=8cm,height=6cm}
}}
\caption{Average $C_{2}$ values for $KB3$ and $KB3p\_f$ interactions in
$^{44}Ti$.}
\label{<C2> for KB3 and KB3p_f in Ti44}
\end{figure}

Note that the recovery of the leading irrep when the spin-orbit interaction is
turned off is clearly signaled not only through an increase in the absolute
values of the first centroid $\left\langle C_{2}\right\rangle $ but also
through the skewness $\delta C_{2}$. For example, in $^{44}Ti$ with the $KB3$
interaction (spin-orbit interaction turned on) the ground state $J=0$ has
$\left\langle C_{2}\right\rangle =110$ and skewness $\delta C_{2}=33$. This
changes for the $KB3p\_f$ interaction to $\left\langle C_{2}\right\rangle =139$
and a skewness of $\delta C_{2}=-37$, as shown in Fig.\ref{<C2> for KB3 and
KB3p_f in Ti44}. The equivalent of the $^{44}Ti$ graph for the $^{48}Ti$ case
is shown in Fig.\ref {<C2> for KB3 and KB3p_f in Ti48}. As for the $^{44}Ti$
case, the results show the recovery of the $SU(3)$ symmetry in $^{48}Ti$ when
the single-particle spin-orbit interaction is turned off.

We now turn to a discussion of the coherence nature of the yrast states. First
notice that the widths of the distributions as defined by $\Delta
C_{2}=\sqrt{\left\langle \left( C_{2}-\left\langle C_{2}\right\rangle \right)
^{2}\right\rangle}$ are surprisingly unaffected (Fig.\ref{<C2> for KB3 and
KB3p_f in Ti44} and Fig.\ref{<C2> for KB3 and KB3p_f in Ti48}) by turning the
spin-orbit interaction on and off. This effect occurs in all cases studied:
$^{44}Ti,$ $^{46}Ti,$ $^{48}Ti$, and $^{48}Cr$. The more detailed graphs,
Fig.\ref{C2 of SU(3) for KB3 in Ti44} and Fig.\ref{C2 of SU(3) for KB3p_f in
Ti44}, offer an explanation in terms of the fragmentation of the $C_{2}$
distribution. As can be seen from these graphs, the irreps that are presented
in the structure of a given yrast state in the presence of the spin-orbit
interaction (Fig.\ref{C2 of SU(3) for KB3 in Ti44}) remain present, even though
with reduced strength, in the structure of the state when the spin-orbit
interaction is turned off (Fig.\ref{C2 of SU(3) for KB3p_f in Ti44}). As a
consequence, $\Delta C_{2}=\sqrt{\left\langle\left( C_{2}-\left\langle
C_{2}\right\rangle \right) ^{2}\right\rangle }$ which measures the overall
spread of contributing irreps, is more or less independent of the spin-orbit
interaction. One can see a sharp decrease in the width of the distribution only
for high spin states like $J=12$ in the graph for $^{44}Ti$ in Fig.\ref{<C2>
for KB3 and KB3p_f in Ti44}.

\begin{figure}[tbp]
\centerline{\hbox{
\epsfig{figure=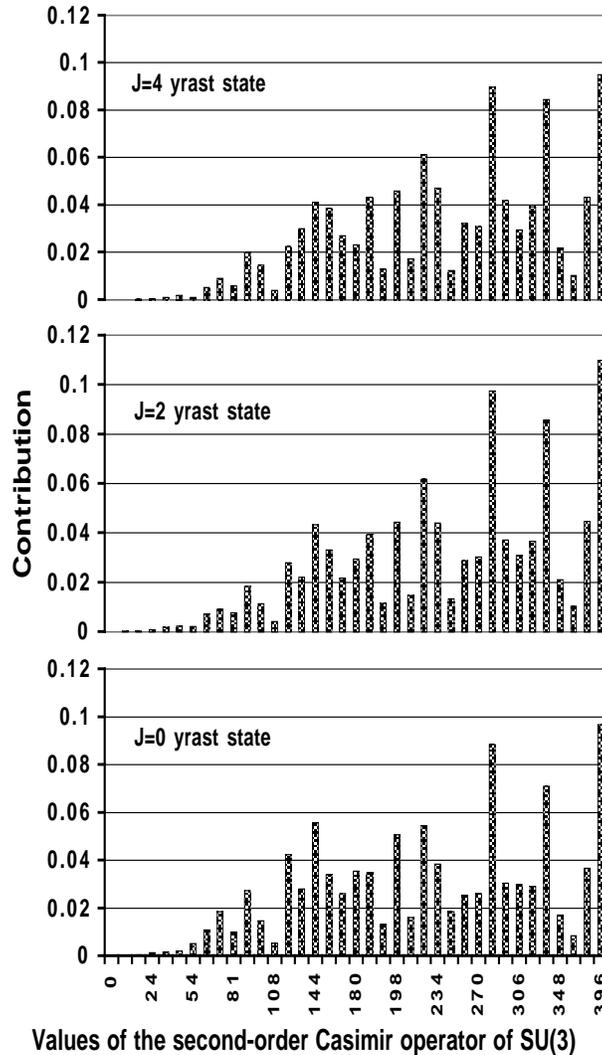,width=8cm,height=14cm}
}}
\caption{Coherent structure of the first three yrast states in
$^{48}Cr$ calculated using realistic single particle energies with Kuo-Brown-3
two body interaction ($KB3$). On the horizontal axis is
$C_{2}$ of $SU(3)$ with contribution of each
$SU(3)$ state to the corresponding yrast state on the vertical axis.}
\label{Coherence in Cr48 yrast band}
\end{figure}

The third type of graph, Fig.\ref{Coherence in Cr48 yrast band}, demonstrates
the coherent nature of the states within the yrast band. The three graphs shown
give the spectrum of the second-order Casimir operator $C_{2}$ of $SU(3)$ for
the $J=0$, 2 and 4 yrast states in $^{48}Cr$. The axes are labelled the same
way as in Figs.\ref{C2 of SU(3) for KB3 in Ti44} and \ref {C2 of SU(3) for
KB3p_f in Ti44}, but in this case all bars are for a single yrast state. In
this figure there are three peaks surrounded by smaller bars that yield a very
similar enveloping shape for the given yrast states. The fragmentation and
spread of $C_{2}$ values is nearly identical for these states with no dominant
irrep, indicative of severe $SU(3)$ symmetry breaking.

Graphs for the $KB3p\_f$ case, when the spin-orbit interaction is turned off,
are not shown since the results are similar to the results for $^{44}Ti$ shown
in Fig.\ref{C2 of SU(3) for KB3p_f in Ti44}. For example, when the spin-orbit
interaction is on (KB3) the leading irrep for $^{48}Cr$ has a $C_{2}$ value of
396 and this account for only around 10$\%$ of the total strength distribution
(see Fig.\ref {Coherence in Cr48 yrast band}), but when the spin-orbit
interaction is off (KB3$p\_f$) the leading irrep is the dominant irrep with
more than 55\% of the total strength.

\begin{figure}[tbp]
\centerline{\hbox{
\epsfig{figure=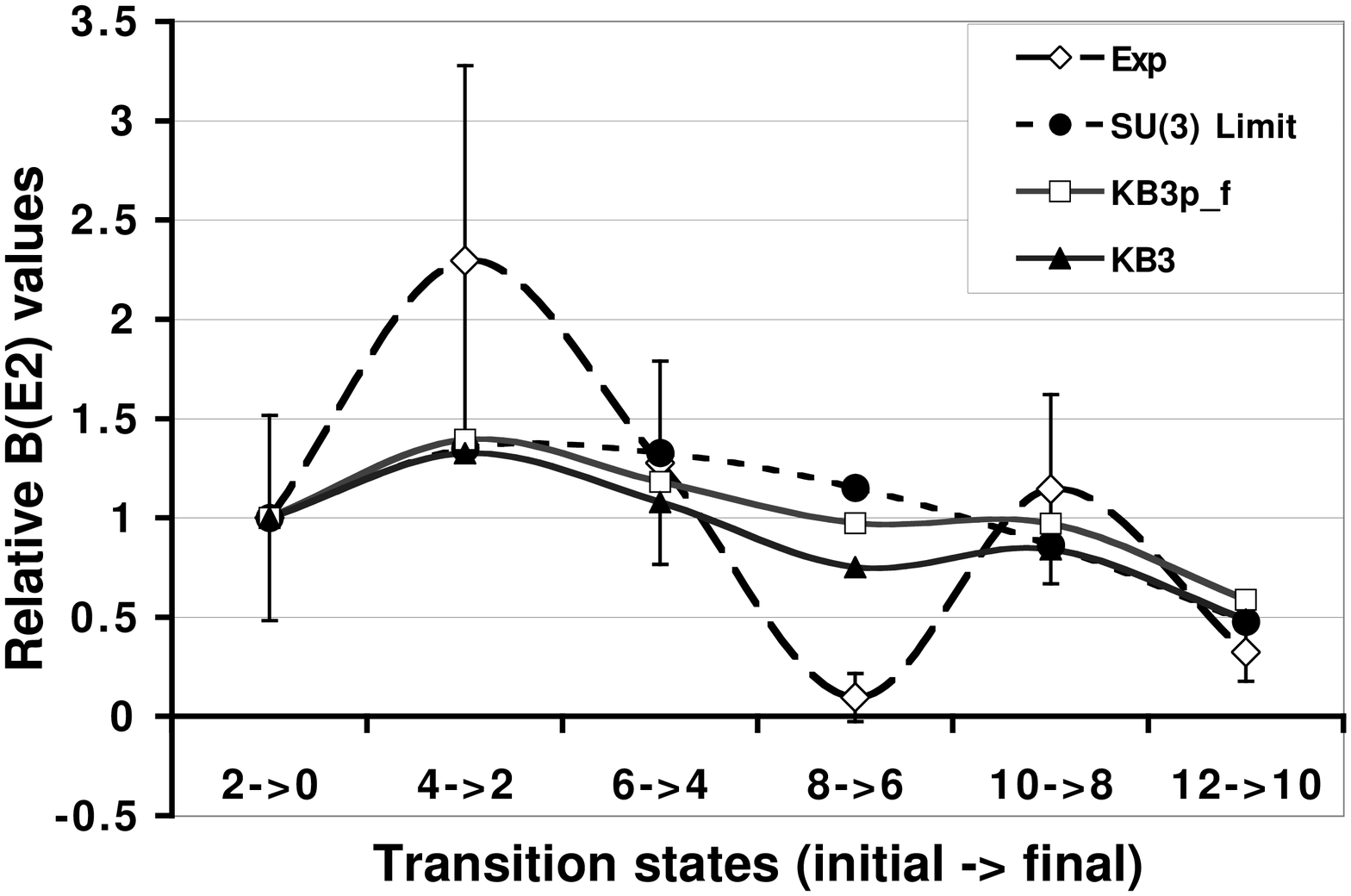,width=8cm,height=6cm}
}}
\caption{Relative $B(E2)$ values $\left( \frac{B(E2:J_{i}\rightarrow J_{f})}{
B(E2:2^{+}\rightarrow 0^{+})}\right) $ for $^{44}Ti.$ The
$B(E2:2^{+}\rightarrow 0^{+})$ transition values are 122.69$e^{2}fm^{4}$ from
experiment, 104.82$e^{2}fm^{4}$ for the KB3 interaction, and 138.58$
e^{2}fm^{4}$ for the $KB3p\_f$ case.}
\label{Relative B(E2) in Ti44}
\end{figure}

\begin{figure}[tbp]
\centerline{\hbox{
\epsfig{figure=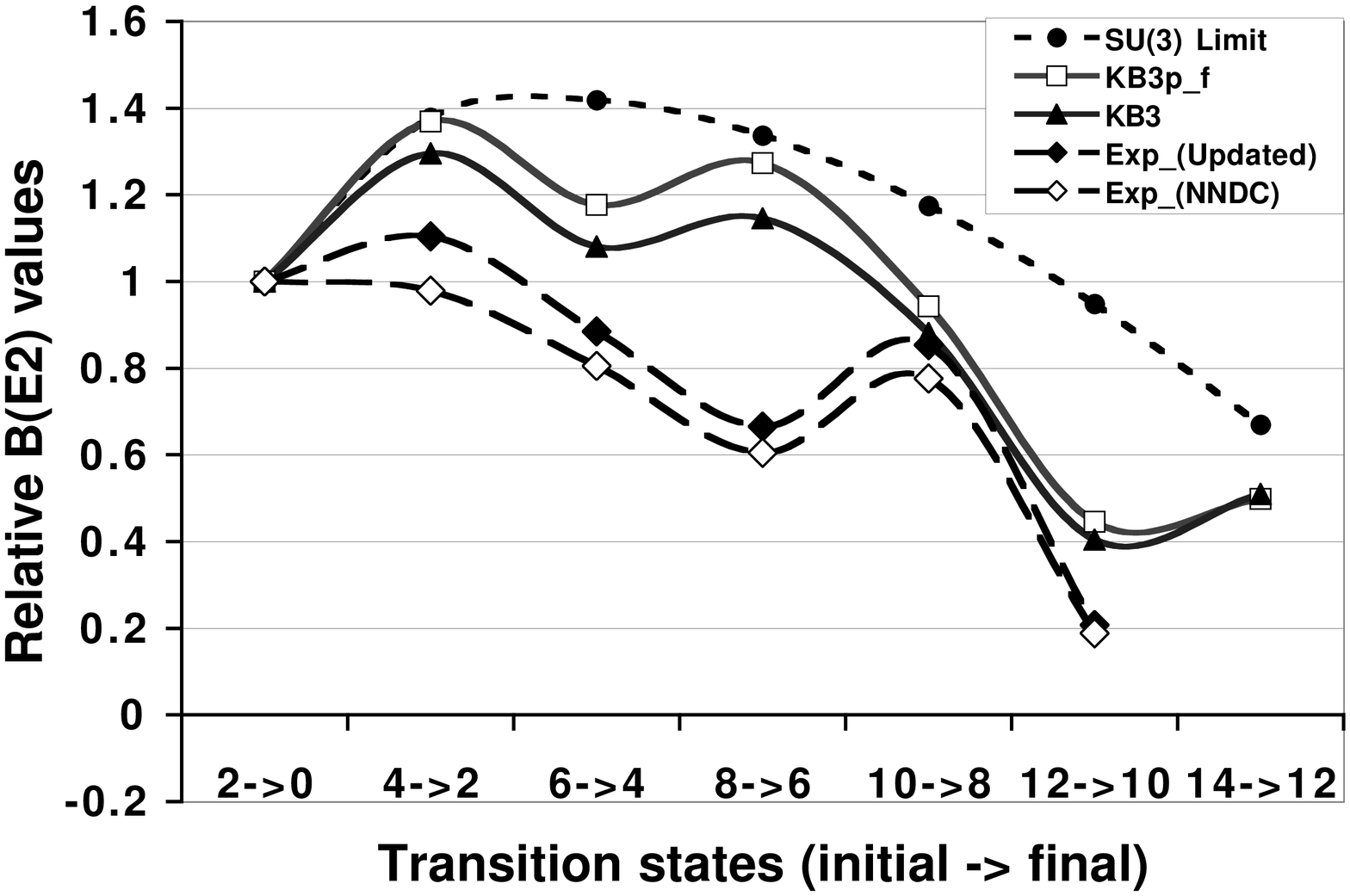,width=8cm,height=6cm}
}}
\caption{Relative $B(E2)$ values $\left( \frac{B(E2:J_{i}\rightarrow J_{f})}{
B(E2:2^{+}\rightarrow 0^{+})}\right) $ for $^{46}Ti.$ The
$B(E2:2^{+}\rightarrow 0^{+})$ transition values are 199.82$e^{2}fm^{4}$ from
experimental data, 181.79$e^{2}fm^{4}$ from updated experimental data,
208$e^{2}fm^{4}$ for KB3 interaction, and 299.83$ e^{2}fm^{4}$ for $KB3p\_f.$}
\label{Relative B(E2) in Ti46}
\end{figure}

\begin{figure}[tbp]
\centerline{\hbox{
\epsfig{figure=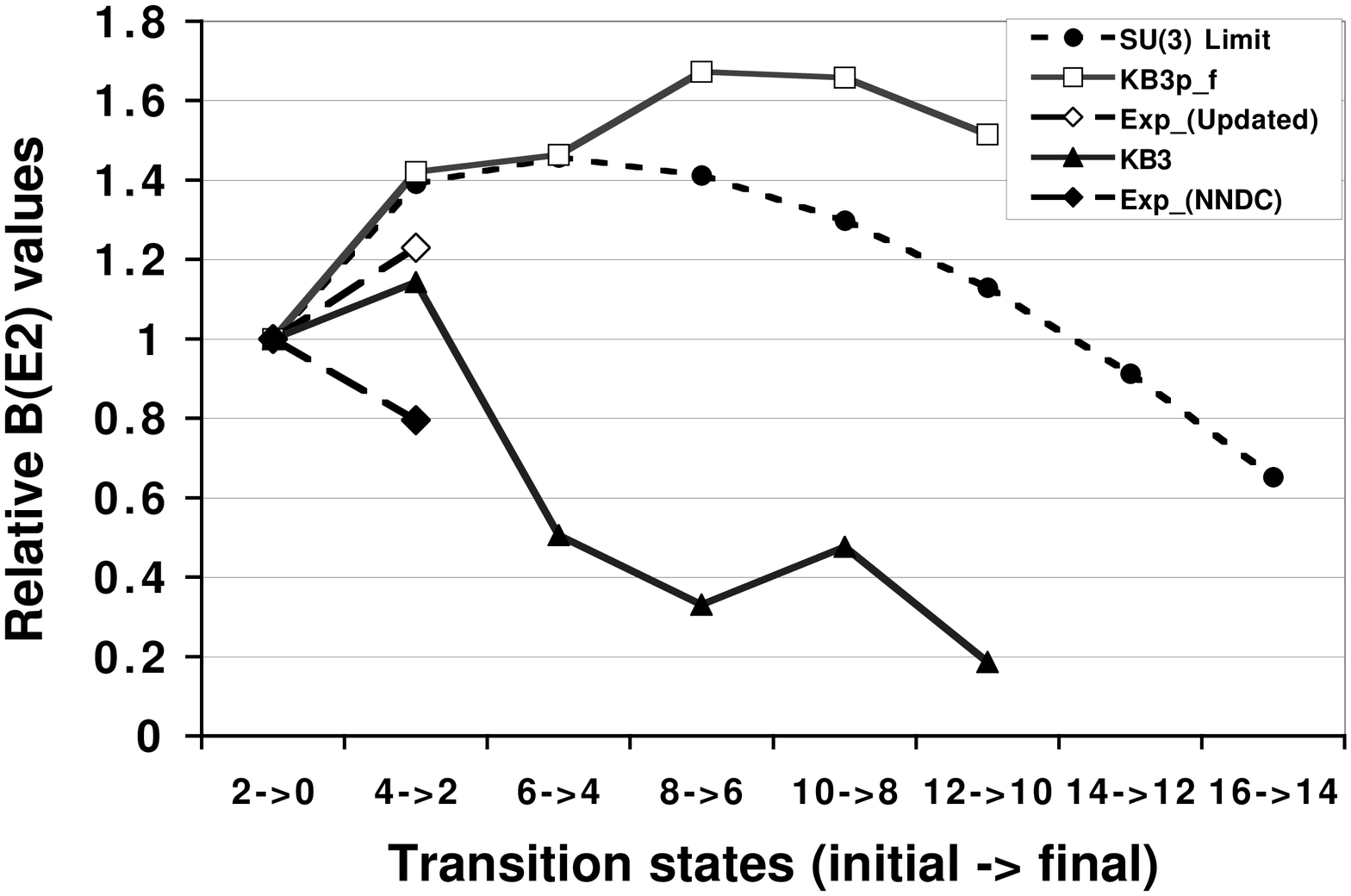,width=8cm,height=6cm}
}}
\caption{Relative $B(E2)$ values $\left( \frac{B(E2:J_{i}\rightarrow J_{f})}{
B(E2:2^{+}\rightarrow 0^{+})}\right) $ for $^{48}Ti.$ The
$B(E2:2^{+}\rightarrow 0^{+})$ transition values are 144.23$e^{2}fm^{4}$ from
experimental data, 155.5$e^{2}fm^{4}$ from updated experimental data,
202.4$e^{2}fm^{4}$ for KB3 interaction, and 445.32$ e^{2}fm^{4}$ for $KB3p\_f.$}
\label{Relative B(E2) in Ti48}
\end{figure}

The last type of graph, Figs.\ref{Relative B(E2) in Ti44}, \ref
{Relative B(E2) in Ti46}, and \ref{Relative B(E2) in Ti48} shows relative
$B(E2)$ values, that is, $B(E2)$ strengths normalized to the
$B(E2:2^{+}\rightarrow 0^{+})$ value. For isoscalar transitions the relative
$B(E2)$ strengths are insensitive to the effective charges which may be used to
bring the theoretical $B(E2:2^{+}\rightarrow 0^{+})$ numbers into agreement
with the experimental values. Whenever an absolute $B(E2:2^{+}\rightarrow
0^{+})$ values are given they are in $e^{2}fm^{4}$ units and the effective
charges are $1.5e$ for protons and $0.5e$ for neutrons ($q_{eff}$ = 0.5).

The first graph on relative $B(E2)$ values ( Fig.\ref{Relative B(E2) in Ti44} )
recaps our results for $^{44}Ti$. Calculated relative $B(E2)$ values for
$^{44}Ti$ corresponding to the spin-orbit interaction turned on (KB3) and
spin-orbit interaction off (KB3$p\_f$) are very close to the pure $SU(3)$
limit. The agreement with experiment is very satisfactory except for the
$4^{+}\rightarrow 2^{+}$ and $8^{+}\rightarrow 6^{+}$ transitions. However, the
experimental data \cite{NNDC} on $8^{+}\rightarrow 6^{+}$ transition gives only
an upper limit of 0.5 pico-seconds to the half-life. We have used the worse
case, namely a half-life of 0.5 ps, as a smaller value would increase the
relative $ B(E2)$ value. For example a half-life of 0.05 ps will agree well
with the relative $B(E2)$ value for the $KB3p\_f$ interaction. This example
supports the adiabatic mixing which seems to be present for all the yrast
states of $^{44}Ti.$

Fig.\ref{Relative B(E2) in Ti46} shows $B(E2)$ values for $^{46}Ti$. In this
case there are deviations from adiabatic mixing for the $6^{+}\rightarrow
4^{+}$, $10^{+}\rightarrow 8^{+}$, and higher transitions. Two experimental
data sets are shown in Fig.\ref{Relative B(E2) in Ti46}: data from the NNDC is
denoted as Exp\_(NNDC), and updated data on $2^{+}\rightarrow 0^{+}$ and 
$4^{+}\rightarrow 2^{+}$ transitions from \cite{Resent Data on B(E2)} is
denoted as Exp\_(Updated). For $^{46}Ti$ the agreement with the experiment is
not as good as for $^{44}Ti$, however the experimental situation is also less
certain. However, the coherent structure is well demonstrated for the first
three yrast states $0^{+},$ $2^{+}$, and $4^{+}$ via relative $B(E2)$ values
for the KB3 and $KB3p\_f$ interactions which are very close to the $SU\left(
3\right) $ limit. 

We conclude this section by showing the recovery of the $SU(3)$ symmetry; this
time via relative $B(E2)$ values as shown for $^{48}Ti$ in Fig.\ref{Relative
B(E2) in Ti48}. In Fig.\ref{Relative B(E2) in Ti48} we see that for the
degenerate single particles case (KB3$p\_f$) the first few transitions have
relative $B(E2)$ values which follow the $SU(3)$ limit very closely. On other
hand, the interaction involving spin-orbit splitting (KB3) is far from the
$SU(3)$ limit. The $B(E2:4^{+}\rightarrow 2^{+})$ transition is strongly
enhanced due to the adiabatic mixing which is missing in the higher than $J=4$
yrast states.

\section{Conclusion and discussion}

The results reported in this paper show that the single-particle spin-orbit
splitting is the primary interaction responsible for breaking of the $SU(3)$
symmetry for nuclei in the lower $fp$-shell. When the spin-orbit splitting is
reduced, as in the $KB3p\_f$ case, the importance of $SU(3)$ as seen through
the dominance of the leading irrep represented in each yrast state is revealed.
It is important to note in this regard that the $p$- and $f$-shells are nearly
degenerate, which implies a small $l^{2}$ splitting.

Although the $SU(3)$ structure of the states is lost in the lower $fp$-shell,
the results also show the mixing of $SU(3)$ irreps that occurs displays
enhanced $B(E2)$ strengths. This adiabatic mixing results in a coherent
structure that is represented in all yrast states for the $^{44}Ti$ case, while
for the other nuclei studied this coherence breaks down after the first few
yrast states. In particular, even though the yrast states are not dominated by
a single $SU(3)$ irrep, the $B(E2:4^{+}\rightarrow 2^{+})$ values remain
strongly enhanced with values close (usually within 10-20\%) to the $SU(3)$
symmetry limit.

\vskip .5cm

Support provided by the Department of Energy under Grant No. DE-FG02-96ER40985
and by the National Science Foundation under Grant No. PHY-9970769 and
Cooperative Agreement EPS-9720652 that includes matching from the Louisiana
Board of Regents Support Fund. The authors also wish to acknowledge the
[Department of Energy's] Institute for Nuclear Theory at the University of
Washington for its hospitality during the final stage of completion of this
work.

\end{document}